\documentclass[journal]{IEEEtran}
\newcommand{\field}[1]{\mathbb{#1}} 

\newcommand{\trace}[1]{\textrm{\textbf{tr}} \left ( {#1} \right )}

\newcommand{\diag}[1]{\textrm{\textbf{ diag}} \left ( {#1} \right )}

\newcommand{\pderiv}[2]{\frac{\partial {#1}}{\partial {#2}} }

\newcommand{\ALGCMT}[1]{$\quad \backslash \backslash$ {#1}}
\newcommand{\BRAK}[1]{\left [ {#1} \right ]}
\newcommand{\CBRAK}[1]{\left \{ {#1} \right \} }
\newcommand{\ABS}[1]{\left | {#1} \right | }
\newcommand{\PAREN}[1]{\left ( {#1} \right )}

\newcommand{\NORM}[1]{\left \| {#1} \right \| }

\newcommand{\FOURIER}[1]{\mathcal{F} \CBRAK{{#1}} }
\newcommand{\IFOURIER}[1]{\mathcal{F}^{-1} \CBRAK{{#1}}  }

\newcommand{\TOP}[1]{\textrm{\textbf{Top}} \left ( {#1} \right ) }

\ifCLASSINFOpdf
\else
\fi

\usepackage[cmex10]{amsmath}
\usepackage{url} 
\usepackage{graphicx, epsfig}
\usepackage{amsfonts}
\usepackage{algorithm, algorithmic}

\begin{document}
%
\title{Sparse Head-Related Impulse Response for Efficient Direct Convolution}
%
%
%

\author{
        Yuancheng Luo, ~\IEEEmembership{Member,~IEEE}, Dmitry N. Zotkin, Ramani Duraiswami, ~\IEEEmembership{Member,~IEEE}
        \thanks{Yuancheng Luo, Dmitry N. Zotkin, and Ramani Duraiswami are with the Perceptual Interfaces and Reality Lab at the University of Maryland Institute for Advanced Computer Studies in College Park, 20742 USA, e-mail: yluo1@umd.edu, dz@umiacs.umd.edu, ramani@umiacs.umd.edu.}\\ 
}

\markboth{IEEE Transactions on Audio, Speech and Language Processing, January~2014}%
{Shell \MakeLowercase{\textit{et al.}}: Bare Demo of IEEEtran.cls for Journals}
%



\maketitle

\begin{abstract}

Head-related impulse responses (HRIRs) are subject-dependent and direction-dependent filters used in spatial audio synthesis. They describe the scattering response of the head, torso, and pinnae of the subject. We propose a structural factorization of the HRIRs into a product of non-negative and Toeplitz matrices; the factorization is based on a novel extension of a non-negative matrix factorization algorithm. As a result, the HRIR becomes expressible as a convolution between a direction-independent \emph{resonance} filter and a direction-dependent \emph{reflection} filter. Further, the reflection filter can be made \emph{sparse} with minimal HRIR distortion. The described factorization is shown to be applicable to the arbitrary source signal case and allows one to employ time-domain convolution at a computational cost lower than using convolution in the frequency domain.
\end{abstract}


%
\IEEEpeerreviewmaketitle

\section{Introduction}
\label{SEC:NMF:INTRO}

The human sound localization ability is rooted in subconscious processing of spectral acoustic cues that arise due to sound scattering off the listener's own anatomy. Such scattering is quantified by a linear, time-invariant, direction-dependent filter known as the Head-Related Transfer Function (HRTF) \cite{BEGAULT}. HRTF knowledge allows presentation of realistic virtual audio sources in a Virtual Auditory Display (VAD) system so that the listener perceives the sound source as external to him/her and positioned at a specific location in space, even though the sound is actually delivered via headphones. A number of additional effects such as environmental modeling and motion tracking are commonly incorporated in VAD for realistic experience \cite{CHENG1999,ZOTKIN2004}.

The HRTF is typically measured by a placing a small microphone in an individual's ear canal and making a recording of a broadband test signal\footnote{Various test signals, such as impulse, white noise, ML sequence, Golay code, frequency sweep, or any broadband signal with sufficient energy in the frequencies of interest can be used for the measurements.} emitted from a loudspeaker positioned sequentially at a number of points in space. The HRTF is the ratio of the spectra of microphone recording at the eardrum and at the head's center position in the absence of the individual. Thus, the HRTF is independent of the test signal and the recording environment and describes the acoustic characteristics of the subject's anthropometry (head, torso, outer ears, and ear canal). The inverse Fourier transform of HRTF is the (time domain) filter's impulse response, called the Head-Related Impulse Response (HRIR).

The primary goal of the current work is to find a short and sparse HRIR representation so as to allow for computationally efficient, low latency time-domain convolution between arbitrary (long) source signal $y$ and short HRIR $x$ \cite{CLARK,BURRUS}. It is expected that direct convolution\footnote{$(x*y)_i = \sum_{j} x_j y_{i-j+1}$ for $x$ and $y$ zero-padded as appropriate} with short and sparse $x$ would be more efficient w.r.t. latency and cost than frequency-domain convolution using the fast Fourier transform (FFT)\footnote{Fourier Transform convolution $x*y = \IFOURIER{\FOURIER{x} \circ \FOURIER{y}}$ for Fourier transform operator $\FOURIER{}$ and element-wise product $\circ$.} \cite{COOLEY,SMITH}. 

Somewhat similar approaches has been explored in the literature previously. In the frequency domain, the HRTF has been decomposed into a product of a common transfer function (CTF) and a directional transfer function (DTF) \cite{CHENG1999,MIDDLEBROOKS,KISTLER}, where the CTF is the minimum-phase filter with magnitude equal to average HRTF magnitude and the DTF is a residual. A more recent work on Pinna-Related Transfer Function (PRTF) \cite{BATTEAU,ALGAZI3,GERONAZZO,RAYKAR} provided successful PRTF synthesis model based on deconvolution of the overall response into \emph{ear-resonance} (derived from the spectral envelope) and \emph{ear-reflection} (derived from estimated spectral notches) parts. The novelty of the current work is that the \emph{time-domain} modeling is considered and constraints are placed on "residual impulse response" (the time-domain analog of the DTF) to allow for fast and efficient real-time signal processing in time domain. Further, the tools to achieve this decomposition (semi-non-negative matrix factorization with Toeplitz constraints) are novel as well.

\section{Problem Formulation}

We propose the following time-domain representation of an HRIR $x \in \field{R}^{M}$ given by
\begin{equation}
\displaystyle
\begin{split}
x \approx f * g, \quad g \geq 0,
\end{split}
\label{EQ:NMF:INTRO:XFG}
\end{equation}
where $*$ is the linear convolution operation, $f \in \field{R}^{M-K+1}$ is a ``common impulse response'' derived from the subject's HRIR set, and $g \in \field{R}^{K}$ is a sparse non-negative ``residual''; the length of $g$ is $K$. In analogy with terms commonly used in PRTF research, hereafter $f$ is called the ``resonance filter'' and $g$ the ``reflection filter''. The resonance filter is postulated to be independent of measurement direction (but of course is different for different subjects), and the directional variability is represented in $g$, which is proposed to represent instantaneous reflections of the source acoustic wave off the listener's anatomy; hence, $g$ is non-negative and sparse. The computational advantage of such a representation is the ability to perform efficient convolution with an arbitrary source signal $y$ via the associative and commutative properties of the convolution operation given by
\begin{equation}
\displaystyle
\begin{split}
y * x = \PAREN{y * f} * g = \PAREN{y * g} * f.
\end{split}
\end{equation}
If $y$ is known in advance, the convolution with $f$ is direction-independent and can be precomputed in advance. Thereafter, direct time-domain convolution with a short and sparse $g$ is fast and can be performed in real time. Moreover, even in the case of streaming $y$, computational savings are possible if the output signal has to be computed for more than one direction (as it is normally the case in VAD for trajectory interpolation). 

\begin{figure}[ht]
  \centering
  \includegraphics[width=.49\textwidth]{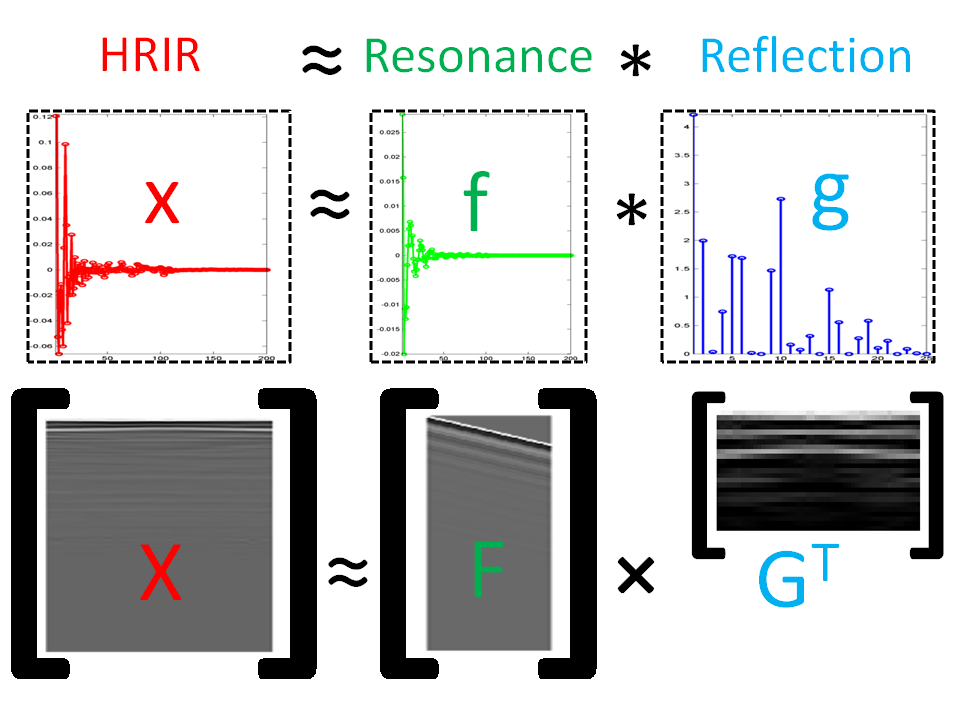}
\caption{Modified semi-non-negative matrix factorization generalizes time-domain convolution for a collection of HRTFs $X$, resonance filter $f$, and non-negative reflection filters in $G$.}
\label{FIG:NMF:INTRO:IRFIG}
\end{figure}

To learn the filters $f$ and $g$, we propose a novel extension of the semi-non-negative matrix factorization (semi-NMF) method \cite{DING}. Semi-NMF factorizes a mixed-signed matrix $X \approx FG^T \in \field{R}^{M \times N}$ into a product of a mixed-signed matrix $F$ and a non-negative matrix $G$ minimizing the approximation error in the least-squares sense. We modify the algorithm so that the matrix $F$ has \emph{Toeplitz structure}; then, $FG^T$ is nothing but a convolution operation with multiple, time-shifted copies of $f$ placed in columns of $F$ (see Fig. \ref{FIG:NMF:INTRO:IRFIG}). Thus, the overall approach for computing $f$ and $g$ is as follows: a) form matrix $X$ from individual HRIRs, placing them as columns; b) run Toeplitz-constrained semi-NMF on $X$; c) take the first column and row of $F$ as $f$; and d) for each direction, obtain non-negative $g$ by taking a corresponding row of $G$.

The paper is organized as follows. In section \ref{SEC:NMF:SNTMF}, the modified semi-NMF algorithm is derived, with further extension to enforce a sparseness constraint on $G$ by formulating it as a regularized $L_1$ norm non-negative least squares problem ($L_1$-NNLS) \cite{LAWSON}. As the cost of time-domain convolution is proportional to the number of non-zero (NZ) elements in $g$, decreasing $K$ (i.e., increasing sparsity) reduces computational load at the cost of increased approximation error. Experimental results are presented in section \ref{SEC:NMF:RESULTS} along with the discussion. Finally, section \ref{SEC:NMF:CONC} concludes the paper.

\section{Semi-non-negative Toeplitz Matrix Factorization}
\label{SEC:NMF:SNTMF}

\subsection{Background}
\label{SEC:NMF:SNTMF:BACKGROUND}

The original non-negative matrix factorization (NMF) \cite{SEUNG} was introduced in the statistics and machine learning literature as a way to analyze a collection of non-negative inputs $X$ in terms of non-negative matrices $F$ and $G$ where $X \approx FG^T$. The non-negativity constraints have been used to apply the factorization to derive novel algorithms for spectral clustering of multimedia data \cite{DING2}. Semi-NMF \cite{DING} is a relaxation of the original NMF where the input matrix $X$ and filter matrix $F$ have mixed sign whereas the elements of $G$ are constrained to be non-negative. Formally, the input matrix $X \in \field{R}^{M \times N}$ is factorized into matrix $F \in \field{R}^{M \times K}$ and matrix $G \in \field{R}^{N \times K}$ by minimizing the residual Frobenius norm cost function
\begin{equation}\label{EQ:NMF:SNTMF:DEF}
\displaystyle
\begin{split}
\min_{F,G} \NORM{X-FG^T}^2_F  & =  \trace{ (X-FG^T)^T (X-FG^T)},
\end{split}
\end{equation}
where $\trace{}$ is the trace operator. For $N$ samples in the data matrix $X$, the $i^{th}$ sample is given by the $M$-dimensional row vector $X_i = X_{:,i}$ and is expressed as the matrix-vector product of $F$ and the $K$-dimensional row vector $G_i = G_{i,:}$. The number of components $K$ is selected beforehand or found via data exploration and is typically much smaller than the input dimension $M$. The matrices $F$ and $G$ are jointly trained using an iterative updating algorithm \cite{DING} that initializes a randomized $G$ and performs an iterative loop computing
\begin{equation}\label{EQ:NMF:SNTMF:UPDATE}
\displaystyle
\begin{split}
F & \leftarrow XG(G^TG)^{-1}, \\
G_{ij} & \leftarrow G_{ij} \sqrt{\frac{(X^TF)^+_{ij} + [G(F^TF)^-]_{ij}}{(X^TF)^-_{ij} + [G(F^TF)^+]_{ij}} }, \\
\PAREN{Q}_{ij}^+ & =  \frac{|Q_{ij}| + Q_{ij}}{2}, \quad \PAREN{Q}_{ij}^- =  \frac{|Q_{ij}| - Q_{ij}}{2}.
\end{split}
\end{equation}
The positive definite matrix $G^TG \in \field{R}^{K \times K}$ in Eq. \ref{EQ:NMF:SNTMF:UPDATE} is small (fast to compute) and the entry-wise \emph{multiplicative updates} for $G$ ensure that it stays non-negative. The method converges to the optimal solution that satisfies \emph{Karush-Kuhn-Tucker} conditions \cite{DING} as the update to $G$ monotonically decrease the residual in the cost function in Eq. \ref{EQ:NMF:SNTMF:DEF} for a fixed $F$, and the update to $F$ gives the optimal solution for the same cost function for a fixed $G$. 

\subsection{Notational Conventions}
\label{SEC:NMF:SNTMF:NOTA}
To modify semi-NMF for learning the direction-independent $f$ and a set of direction-dependent $g$, we introduce the following notation. Assume that $\tilde{F}$ is a Toeplitz-structured matrix and $\tilde{F}_{ij} = \Theta_{i-j}$ for parameters $\Theta = [\Theta_{1-M}, \hdots, \Theta_{K-1}]^T$; thus, all entries along diagonals and sub-diagonals of $\tilde{F}$ are constant. Hence, the Toeplitz structure is given by
\begin{equation}\label{EQ:NMF:SNTMF:NOTA:TOP}
\displaystyle
\begin{split}
\TOP{\Theta} = \BRAK{\begin{array}{ccccc}
\Theta_0 & \Theta_1 & \hdots & \Theta_{K-2}&  \Theta_{K-1}\\
\Theta_{-1} & \Theta_0 & \Theta_1 & \hdots & \Theta_{K-2}\\
\vdots & \ddots & \ddots & \ddots & \vdots \\
\Theta_{2-M} & \hdots  & \Theta_{-1} & \Theta_0 & \Theta_{1} \\
\Theta_{1-M} & \Theta_{2-M} &  \hdots & \Theta_{-1}& \Theta_0
\end{array}},
\end{split}
\end{equation}
and is fully specified by parameters $\CBRAK{\Theta_0, \hdots, \Theta_{K-1}}$ and $\CBRAK{\Theta_0, \hdots, \Theta_{1-M}}$ along the first row and column. The Toeplitz matrix can also be represented indirectly as a linear combination of the parameters weighted by shift matrices $S^k \in \field{R}^{M \times K}$ as
\begin{equation}\label{EQ:NMF:SNTMF:NOTA:ALP}
\displaystyle
\begin{split}
\tilde{F} = \sum_{k=1-M}^{K-1} S^k \Theta_{k}, \quad S^k_{ij} = \delta_{i,j-k}.
\end{split}
\end{equation}

An arbitrary matrix $F$ can be approximated by its nearest Toeplitz matrix $\tilde{F}$, which is defined as the minimizer of the residual Frobenius norm cost function given by
\begin{equation}\label{EQ:NMF:SNTMF:NOTA:TOPPROB}
\displaystyle
\begin{split}
J & = \NORM{ F - \tilde{F}}^2_F = \trace{F^T F - 2F^T \tilde{F} + \tilde{F}^T \bar{F}},  \\
\pderiv{J}{\Theta_k} & = 2\trace{(F-\tilde{F})^T  \pderiv{\tilde{F}}{\Theta_k} }, \quad  \pderiv{\tilde{F}}{\Theta_k} = S^k, \\
\end{split}
\end{equation}
where the partial derivatives of $J$ w.r.t. $\Theta_k$ are linearly independent due to the trace term. By equating the derivatives to zero, the solution $\Theta$ is given by
\begin{equation}\label{EQ:NMF:SNTMF:NOTA:TOPSOL}
\displaystyle
\begin{split}
\Theta_k & =  \frac{\trace{F^T S^k}}{\min(k+M, K-k, K, M)}.
\end{split}
\end{equation}
Hence, a Toeplitz approximation $\tilde{F}$ to an arbitrary matrix $F$ is obtained simply by taking the means of the subdiagonals of $F$. 

\subsection{Toeplitz-Constrained Semi-NMF}
\label{SEC:NMF:SNTMF:UNIQ}

Assuming that a solution of the factorization problem $F$ has in fact Toeplitz structure as per Eq. \ref{EQ:NMF:SNTMF:NOTA:ALP};  the cost function in Eq. \ref{EQ:NMF:SNTMF:DEF} is quadratic (convex) w.r.t. each $\Theta_k$ and the set of parameters $\Theta$ has a unique minimizer. The partial derivatives of the cost function\footnote{Unlike the case considered in section \ref{SEC:NMF:SNTMF:NOTA}, the partial derivatives in Eq. \ref{EQ:NMF:SNTMF:UNIQ:TOP} are linearly dependent.} are given by
\begin{equation}\label{EQ:NMF:SNTMF:UNIQ:TOP}
\displaystyle
\begin{split}
 \pderiv{\NORM{X-\tilde{F}G^T}^2_F}{\Theta_k}  = \pderiv{\trace{ (X-\tilde{F}G^T)^T (X-\tilde{F}G^T)}}{\Theta_k}  \\ 
 =  2\trace{ \PAREN{G^T G \sum_{i=1-K}^{M-1} S^{k^T}  S^{i}  \Theta_{i}}  - S^{k^T}  X G},
\end{split}
\end{equation}
where the product of shift matrices $S^{k^T} S^i$ can be expressed as the square shift matrix $\bar{S}^{i-k}$. To solve for the set of parameters $\Theta$, one needs to set the partial derivatives to zero, which yields a linear equation $A\Theta = b$ where $A \in \field{R}^{|\Theta| \times |\Theta|}$, $|\Theta| = M+K-1$ is a Toeplitz square matrix, and $b \in \field{R}^{M \times 1}$ is a vector specified as
\begin{equation}\label{EQ:NMF:SNTMF:UNIQ:UPDATE}
\displaystyle
\begin{split}
A_{M+k,M+i} = \trace{G^T G \bar{S}^{i-k} }, \quad b_{M+k} = \trace{S^{k^T} XG}.
\end{split}
\end{equation}
For positive-definite $A$, the matrix $\tilde{F}$ is given by the linear equation solution:
\begin{equation}\label{EQ:NMF:SNTMF:UNIQ:TOPSOL}
\displaystyle
\begin{split}
\tilde{F} = \TOP{\Theta}, \quad \Theta = A^{-1}  b,
\end{split}
\end{equation}
which is the unique minimizer of Eq. \ref{EQ:NMF:SNTMF:DEF}. Thus, to enforce Toeplitz structure on $F$, the iterative update $F \leftarrow XG(G^TG)^{-1}$ in the semi-NMF algorithm (Eq. \ref{EQ:NMF:SNTMF:UPDATE}) is replaced by computing $F$ as prescribed by Eq. \ref{EQ:NMF:SNTMF:UNIQ:UPDATE} and Eq. \ref{EQ:NMF:SNTMF:UNIQ:TOPSOL}.

\begin{figure*}[ht]
  \centering
\includegraphics[width=.85\textwidth]{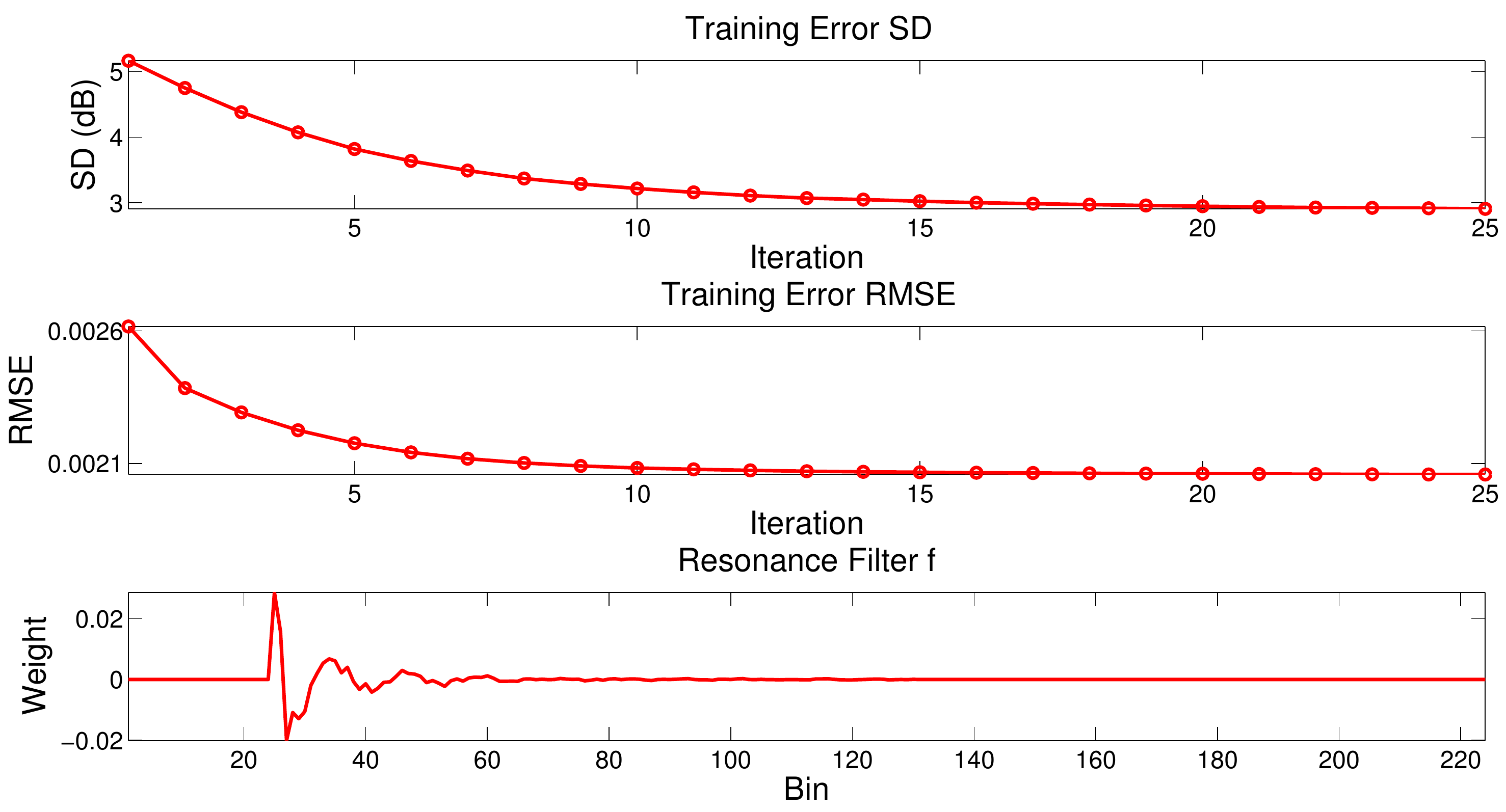}
\caption{RMSE / SD error progress over 25 algorithm iterations.}
\label{FIG:NMF:RESULTS:TRAIN:F}
\end{figure*}

Note that to perform a convolution between $f$ and $g$ (i.e., to reconstruct the HRIR) one needs to further constrain the Toeplitz matrix $\tilde{F}$ given in Eq. \ref{EQ:NMF:SNTMF:NOTA:TOP} in order to fulfill the filter length requirements. Such convolution is equal to the constrained Toeplitz matrix-vector product
\begin{equation}\label{EQ:NMF:SNTMF:UNIQ:TVP}
\displaystyle
\begin{split}
X_i = \BRAK{\begin{array}{cccc}
\Theta_0 & 0 & \hdots & 0\\
\Theta_{-1} & \Theta_0 & 0 & \hdots \\
\vdots & \hdots & \ddots & 0 \\
\Theta_{K-M} &  \hdots & \Theta_{-1}& \Theta_0\\
0 & \Theta_{K-M} & \hdots & \Theta_{-1}\\
\vdots &\hdots &\ddots & \vdots \\
0 & \hdots & 0 & \Theta_{K-M}\\
\end{array}}
\BRAK{\begin{array}{c}
G_{i1} \\
\vdots \\
G_{iK}
\end{array}},
\end{split}
\end{equation}
where the parameters $\CBRAK{\Theta_{K-M-1}, \hdots, \Theta_{1-M}, \Theta_1, \hdots, \Theta_K}$ are set to zero. Only the NZ parameters $\CBRAK{\Theta_0,\hdots \Theta_{K-M}}$ are solved for in a smaller $(M-K+1) \times (M-K+1)$ sized linear system as per Eq. \ref{EQ:NMF:SNTMF:UNIQ:UPDATE} and Eq. \ref{EQ:NMF:SNTMF:UNIQ:TOPSOL}. These NZ parameters form the resonance filter $f$:
\begin{equation}
\label{EQ:NMF:SNTMF:UNIQ:F}
\displaystyle
\begin{split}
f = \CBRAK{\Theta_0,\hdots \Theta_{K-M}} \in \field{R}^{M-K+1}.
\end{split}
\end{equation}

\subsection{Minimizing the Number of Reflections}
\label{SEC:NMF:SNTMF:SPARSE}

To introduce sparsity, we restrict the number of NZ entries (NNZE) in $G$. In order to do that, we fix the trained resonance filter $\tilde{F}$ and solve for each reflection filter $g=G_i$ separately in a penalized $L_1$-NNLS problem formulation \cite{KIM2} given by
\begin{equation}\label{EQ:NMF:SNTMF:SPARSE:L1}
\displaystyle
\begin{split}
\min_{G_i} \NORM{\mathcal{D} \PAREN{FG_i^T-X_i}}_2^2 + \lambda \ABS{G_i}_1, \quad \textrm{s.t. } G_i \geq 0,
\end{split}
\end{equation}
where $D \in \mathcal{R}^{M_* \times M}$ is some transformation of the residual\footnote{A free Matlab solver for $L_1$-NNLS is available online at \url{http://www.stanford.edu/~boyd/papers/l1_ls.html}}. Three transformations are considered.

\begin{figure*}[ht]
  \centering
\includegraphics[width=.49\textwidth]{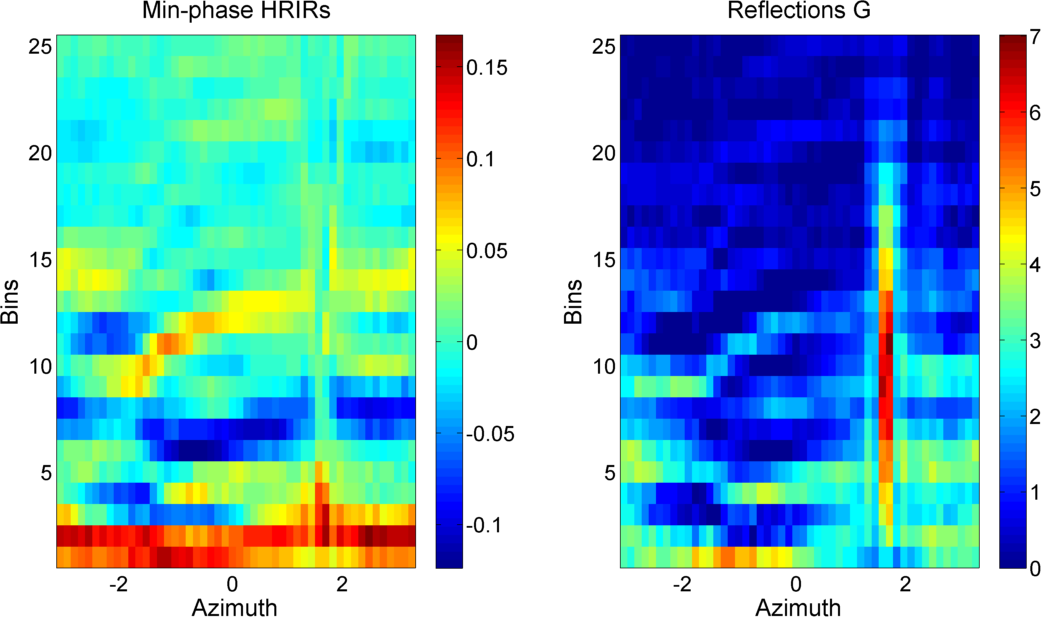} 
\includegraphics[width=.49\textwidth]{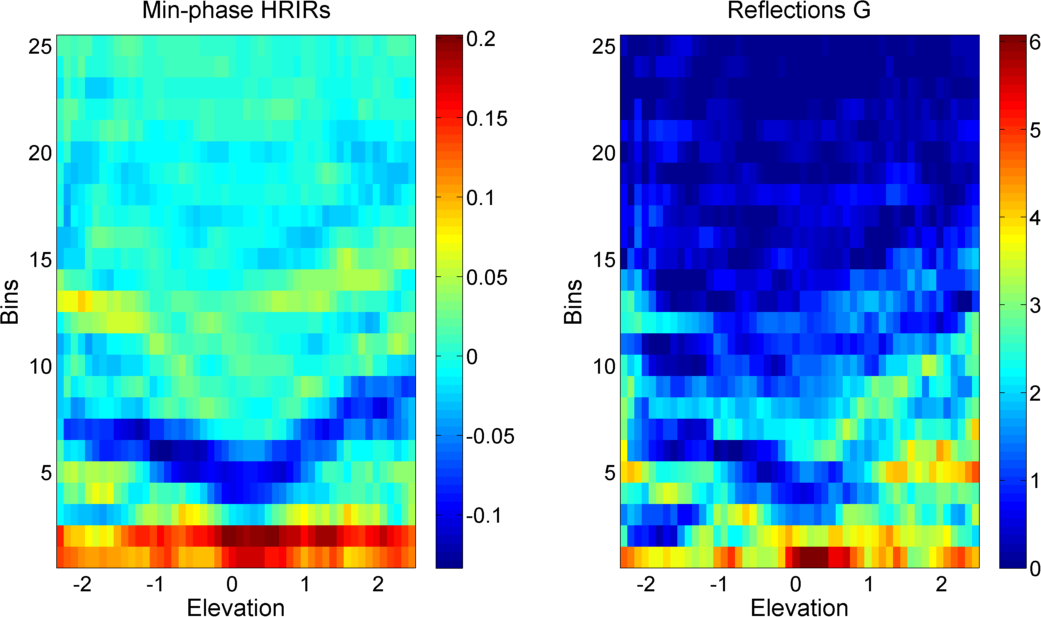} \\ \vspace{.15cm}
\includegraphics[width=.49\textwidth]{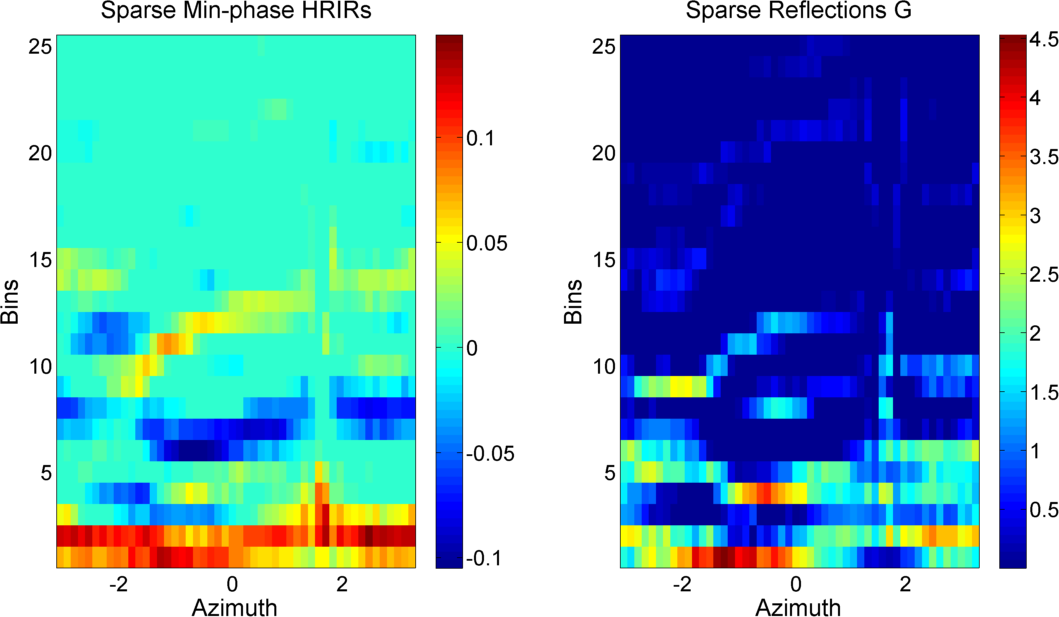} 
\includegraphics[width=.49\textwidth]{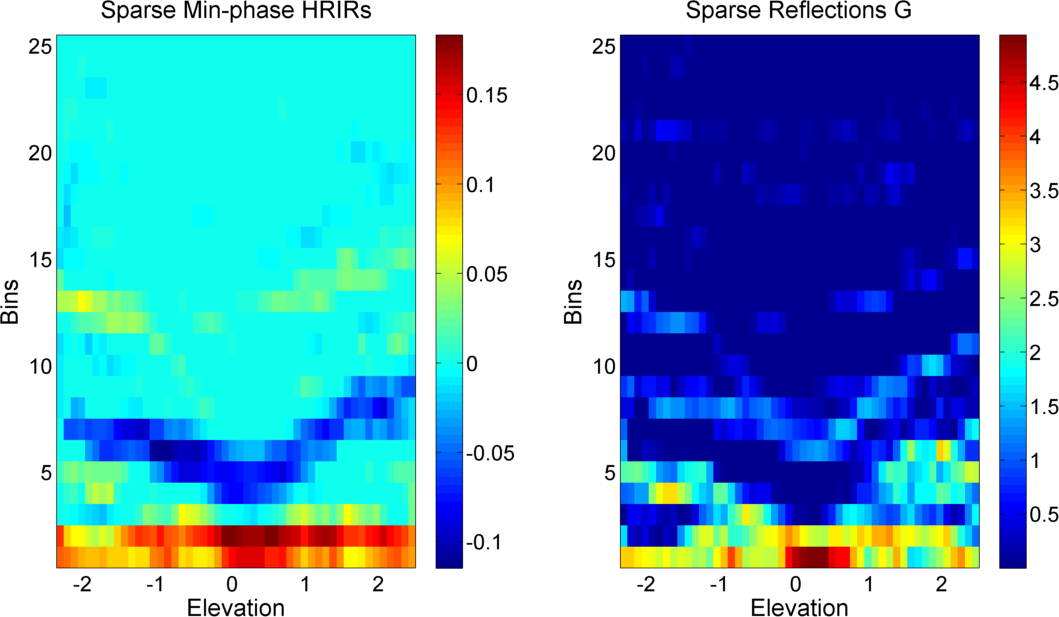} 
\caption{Top row: Slices of reflection filter matrix $G$ trained without sparsity constraint; also, original HRIR after min-phase processing, time delay removing, and normalization. Bottow row: Slices of reflection filter matrix $G$ trained with sparsity constraint applied ($\lambda = 10^{-3}$); also, HRTR reconstructed from it.}
\label{FIG:NMF:RESULTS:TRAIN:NAZELEV}
\end{figure*}

\begin{bf}1.\end{bf} The identity transform $\mathcal{D}_{I} = I \in \field{R}^{M \times M}$, which directly minimizes the residual norm while penalizing large magnitudes in the reflection filter $G_i$.

\begin{bf}2.\end{bf} The convolution transform
\begin{equation}\label{EQ:NMF:SNTMF:SPARSE:DC}
\displaystyle
\begin{split}
\mathcal{D}_{C} & = \TOP{\Theta^C} \in \field{R}^{M \times M}, \\
\Theta^C_{1:M-1} & = \mathcal{N}_{\sigma}(1:M-1), \quad \Theta^C_{0:1-M} = \mathcal{N}_{\sigma}(0:1-M),
\end{split}
\end{equation}
which is characterized by the Gaussian filter $\mathcal{N}_{\sigma} (x) = \frac{1}{\sigma \sqrt{2 \pi}} e^{-\frac{x^2}{2 \sigma^2} }$. This transform effectively low-passes the reconstructed HRIR. It is equivalent\footnote{Convolution in time domain is equivalent to windowing in frequency domain, and vice versa.} to windowing the frequency-domain residuals with a Gaussian filter of inverse bandwidth; hence, the low-frequency bins are weighted heavier in the reconstruction error.

\begin{bf}3.\end{bf} The window transform
\begin{equation}\label{EQ:NMF:SNTMF:SPARSE:DW}
\displaystyle
\begin{split}
\mathcal{D}_{W} = \diag{v_{\sigma}(0:M-1)} \in \field{R}^{M \times M},
\end{split}
\end{equation}
where $v_\sigma(x) = e^{-\frac{x^2}{\sigma^2}}$ is a Gaussian-like filter. The window transform has the effect of convolving the signal spectrum with a filter $v_\sigma(x)$ as if both were time series, which is equivalent to windowing HRIR in time domain by the Gaussian filter of inverse bandwidth. In this way, the earlier parts of the reconstructed HRIR contribute to the reconstruction error to the larger extent.

The additional regularization term $\lambda$ in Eq. \ref{EQ:NMF:SNTMF:SPARSE:L1} affects the sparsity of $g$ as increasing $\lambda$ decreases the NNZE. In our practical implementation, we also discard elements that are technically non-zero but have small ($\leq 10^{-4}$ magnitude) as they contribute little to the reconstruction. The final algorithm for learning the resonance and reflection filters with the sparsity constraint on the latter is summarized in Algorithm \ref{ALG:NMF:TOP}.

\begin{algorithm}
\caption{Modified Semi-NMF for Toeplitz Constraints}
\label{ALG:NMF:TOP}
\begin{algorithmic}[1]
\REQUIRE  Filter length $K$, transformation matrix $D \in \field{R}^{M* \times M}$, HRIR matrix $X \in \field{R}^{M \times N}$, max-iterations $T$
\STATE $G \gets \textbf{rand}(N,K)$ \ALGCMT{Random initialization}
\FOR {$t = 1$ to $T$}
	\STATE $\Theta \gets A^{-1}b$ \ALGCMT{Solve for resonance via Eqs. \ref{EQ:NMF:SNTMF:UNIQ:UPDATE}, \ref{EQ:NMF:SNTMF:UNIQ:TOPSOL}}
	\STATE $\tilde{F} \gets \TOP{\Theta}$ \ALGCMT{Toeplitz matrix via Eqs. \ref{EQ:NMF:SNTMF:UNIQ:TVP}, \ref{EQ:NMF:SNTMF:UNIQ:F}}
	\STATE Update $G$. \ALGCMT{Multiplicative update via Eq. \ref{EQ:NMF:SNTMF:UPDATE}}
\ENDFOR
\STATE Fine-tune $G$. \ALGCMT{Vary $\lambda$, $\sigma$ in Eqs. \ref{EQ:NMF:SNTMF:SPARSE:L1}, \ref{EQ:NMF:SNTMF:SPARSE:DW}, \ref{EQ:NMF:SNTMF:SPARSE:DC}}
\RETURN $\tilde{F}$, $G$
\end{algorithmic}
\end{algorithm}

\section{Results}
\label{SEC:NMF:RESULTS}

\subsection{HRIR/HRTF Data Information}

We have performed an extensive series of experiments on the data from the the well-known CIPIC database \cite{ALGAZI}; however, the approach can be used with arbitrary HRTF data \cite{GARDNER,GUPTA,WARUSFEL,ZOTKIN2006}. We pre-process the data as follows: a) convert HRIR to min-phase; b) remove the initial time delay so that the onset is at time zero; and c) normalize each HRIR so that the absolute sum over all samples is equal to unity.

As mentioned previously, our processing intends to separate the arbitrary impulse response collection of into ``resonance'' (direction-independent) and ``reflective'' (direction-dependent) parts. For the HRIR, we believe that these may correspond to pinna/head resonances and instantaneous reflections off the listener's anthropometry, respectively. Such an approach may also be applicable to other IR collections; for example, room impulse responses \cite{JEUB} may be modeled as a convolution between a shared ``resonance'' filter (i.e. long reverberation tail) and the ``reflective'' filter (early sound reflections off the walls). In order to obtain a unique decomposition using Algorithm \ref{ALG:NMF:TOP}, one would need to have the number of directional IR measurements larger than the IR filter length, which may be impractical. This topic is a subject of future research.

\subsection {Error Metric}
\label{SEC:NMF:RESULTS:EM}

For evaluation, we consider two error metrics -- the root-mean square error (RMSE) and the spectral distortion (SD), representing time-domain and frequency-domain distortions respectively:
\begin{equation} \label{EQ:NMF:RESULTS:ERR}
\displaystyle
\begin{split}
\textrm{RMSE} & = \sqrt{\frac{\NORM{\PAREN{X-\tilde{F}G^T}}^2_F}{MN}}, \\ 
\textrm{SD}\PAREN{H^{\CBRAK{j}}, \tilde{H}^{\CBRAK{j}}} & = \sqrt{\frac{1}{M} \sum_{i=1}^M \PAREN{ 20 \log_{10} \frac{|H^{\CBRAK{j}}_i|}{|\tilde{H}^{\CBRAK{j}}_i|} }^2 },
\end{split}
\end{equation}
where $X_j$ is the reference HRIR, $\tilde{F}G_j^T$ is the reconstruction of it, $H^{\CBRAK{j}} = \FOURIER{X_j}$ is the reference HRTF, $X_j$ is the reference HRIR, and $\tilde{H}^{\CBRAK{j}} = \FOURIER{\tilde{F}G_j^T}$ is the HRTF reconstruction.

Another feasible comparison is validation of the reconstruction derived from sparse representation (Eq. \ref{EQ:NMF:SNTMF:SPARSE:L1}) against the naive regularized least squares ($L_1$-LS) approximation of HRIR $X_i$ given by
\begin{equation}\label{EQ:NMF:SPARSE:L1NOF}
\displaystyle
\begin{split}
\min_{\hat{x}} \NORM{\mathcal{D} \PAREN{ \hat{x}-X_i}}_2^2 + \lambda \ABS{\hat{x}}_1, 
\end{split}
\end{equation}
where $\hat{x} \in \field{R}^{M \times 1}$ (i.e. magnitude-constrained approximation without non-negativity constraint). The difference between SD error of $L_1$-NNLS approximation and of $L_1$-LS approximation is a metric of advantage provided by our algorithm in comparison with LS HRIR representation, which retains large-magnitude HRIR components irrespective of their sign.

\subsection{Resonance and Reflection Filter Training}
\label{SEC:NMF:RESULTS:TRAIN}

The resonance and reflection filters $f$ and $G$ are jointly trained via Algorithm \ref{ALG:NMF:TOP} for $50$ iterations for $N=1250$ number of samples, $M=200$ time-bins, and $K=25$ filter length using left-ear data of CIPIC database subject 003. $N$ and $M$ here are fixed (they are simply the parameters of the input dataset). The choice of $K$ is somewhat arbitrary and should be determined experimentally to obtain the best compromise between computational load and reconstruction quality. Here we set it to the average human head diameter ($\approx 19.2$ cm) at the HRIR sampling frequency ($44100$ Hz). Visual HRIR examination reveals that most of the signal energy is indeed concentrated in the first $25$ signal taps.

Fig. \ref{FIG:NMF:RESULTS:TRAIN:F} shows RMSE and SD error over $50$ iterations of Algorithm \ref{ALG:NMF:TOP} with no sparsity constraint on $G$ (i.e. $\lambda = 0.0$). The final filter $f$ is a periodic, decaying functions resembling a typical HRIR plot. The final matrix $G$ is shown in the top row of Fig. \ref{FIG:NMF:RESULTS:TRAIN:NAZELEV}. The mean NNZE for $G$ is $22.74$ (it is less than $K$ due to removal of all elements with magnitude less than $10^{-4}$). As it can be seen, the SD error achieved is $3.0$ dB over the whole set of directions.

In order to obtain the sparse HRIR representation, we re-ran the algorithm using identity transformation in $L_1$-NNLS constraint and a fixed $\lambda = 10^{-3}$ (this parameter was determined empirically to cut the NNZE approximately in half). The final matrix $G$ obtained in this case is shown in the bottom row of Fig. \ref{FIG:NMF:RESULTS:TRAIN:NAZELEV}. It is sparse as expected and has a number of non-zero bands spanning the time-direction domain; thus, only the most salient components of $G$ are retained. In this case, the mean NNZE is $11.48$ and the SD error is $5.3$ dB over the whole set of directions. In the following section, the guidelines for setting $\lambda$ are considered.

\subsection{Regularization Term Influence}
\label{SEC:NMF:RESULTS:LAMBDA}

\begin{figure*}[ht]
  \centering
  \includegraphics[width=.8\textwidth]{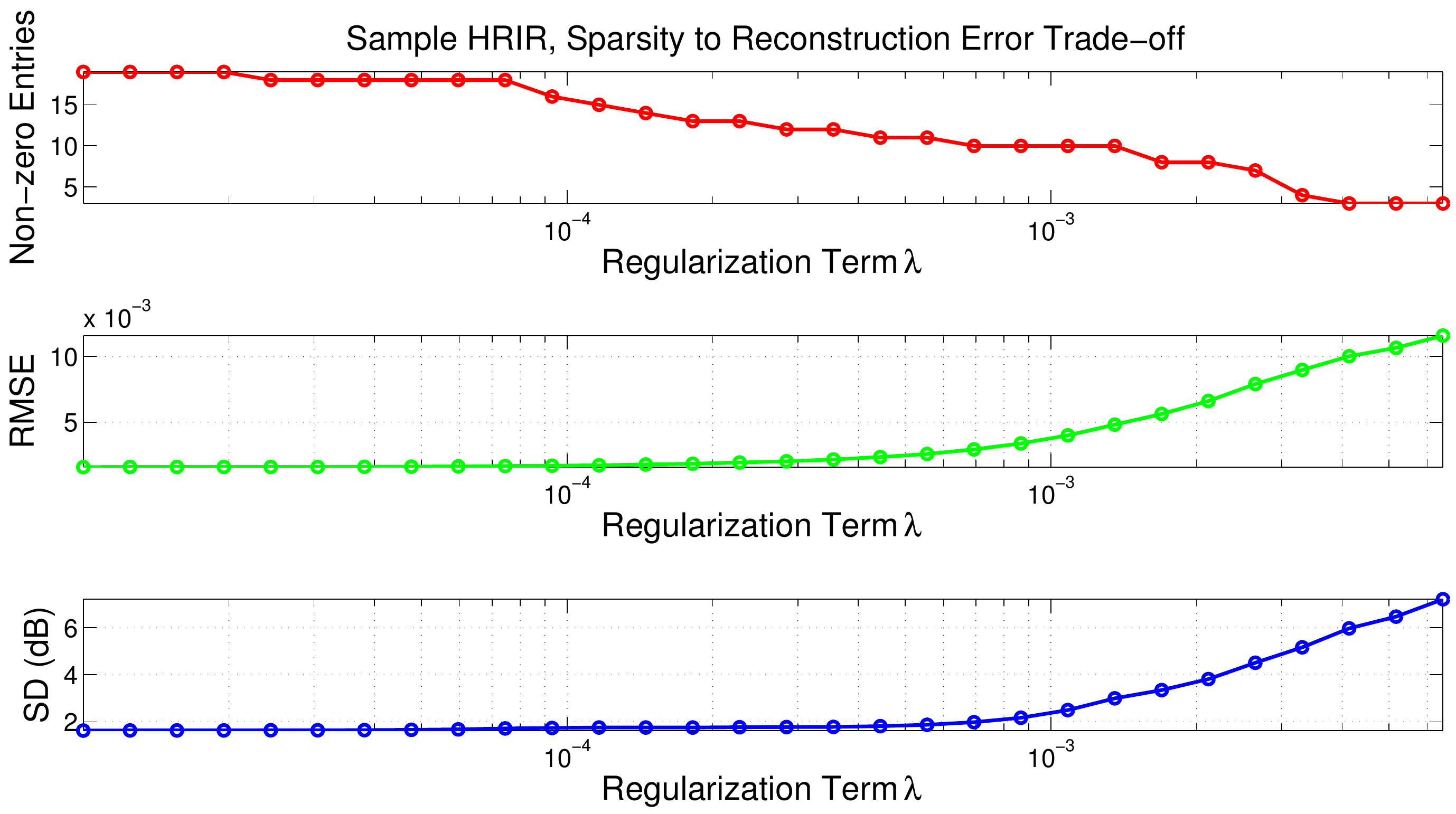}
\includegraphics[width=.8\textwidth]{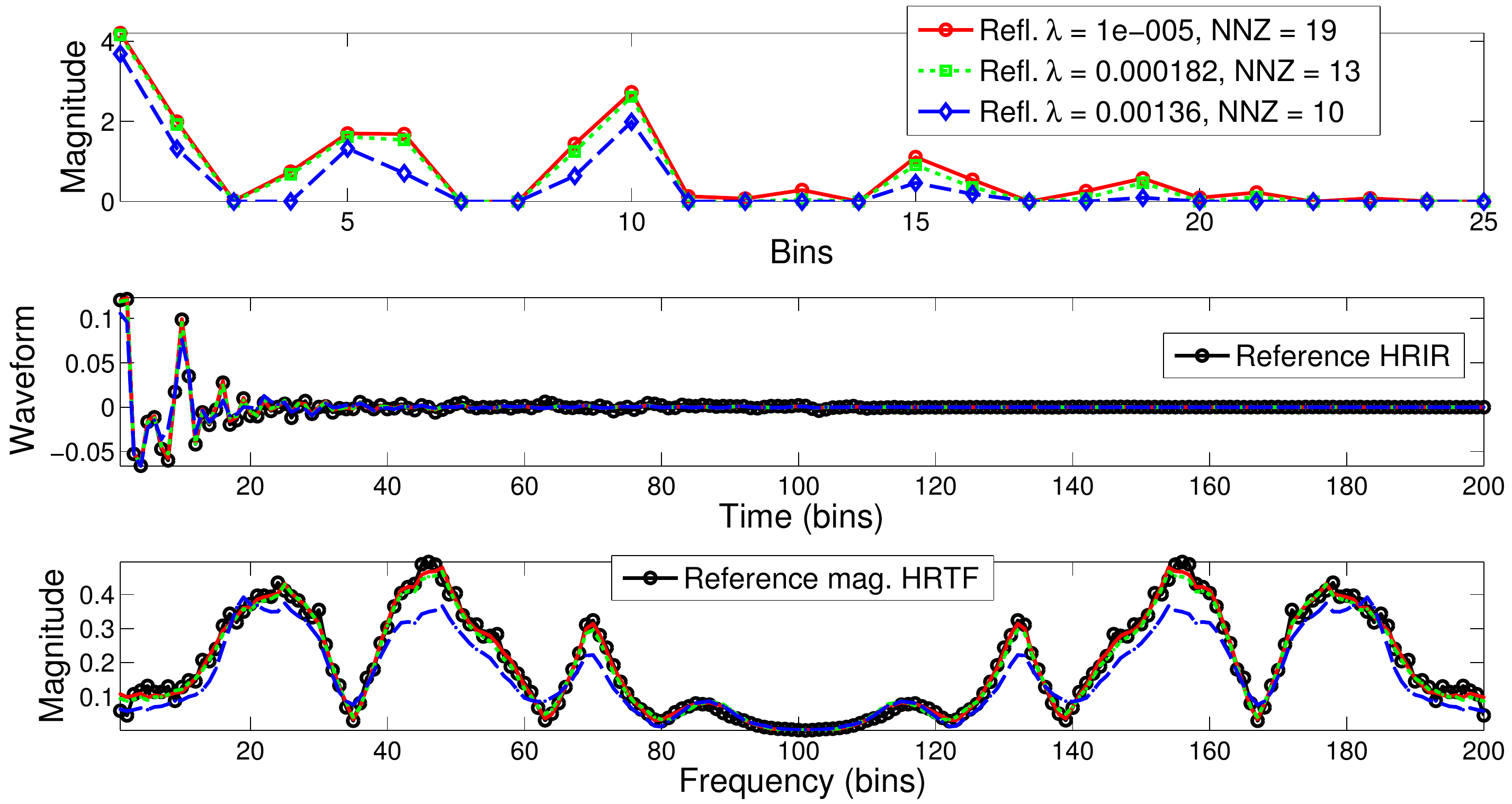}
\caption{Influence of the $L_1$ regularization term $\lambda$ in Eq \ref{EQ:NMF:SNTMF:SPARSE:L1} on NNZE and on the reconstruction error for sample HRIR.}
\label{FIG:NMF:SPARSE:L1}
\end{figure*}

We investigate the effects of varying the $\lambda$ term in Eq. \ref{EQ:NMF:SNTMF:SPARSE:L1} under the identity transform $\mathcal{D}_I$ on the NNZE in $G$ and on the RMSE / SD error. A sample HRIR is chosen randomly from the data set. Fig. \ref{FIG:NMF:SPARSE:L1} shows the effect of changing $\lambda$ on NNZE, RMSE, SD error, and reconstructed HRIR/HRTF \emph{per se}. The trends that one can see in the figure are consistent with expectation; it is interesting to note that as $\lambda$ increases, low-magnitude elements in $G$ are discarded whereas both the dominant time-domain excitations and the shape of the spectral envelope in the reconstructed HRIR are preserved.

\begin{figure}[ht]
  \centering
  \includegraphics[width=.49\textwidth]{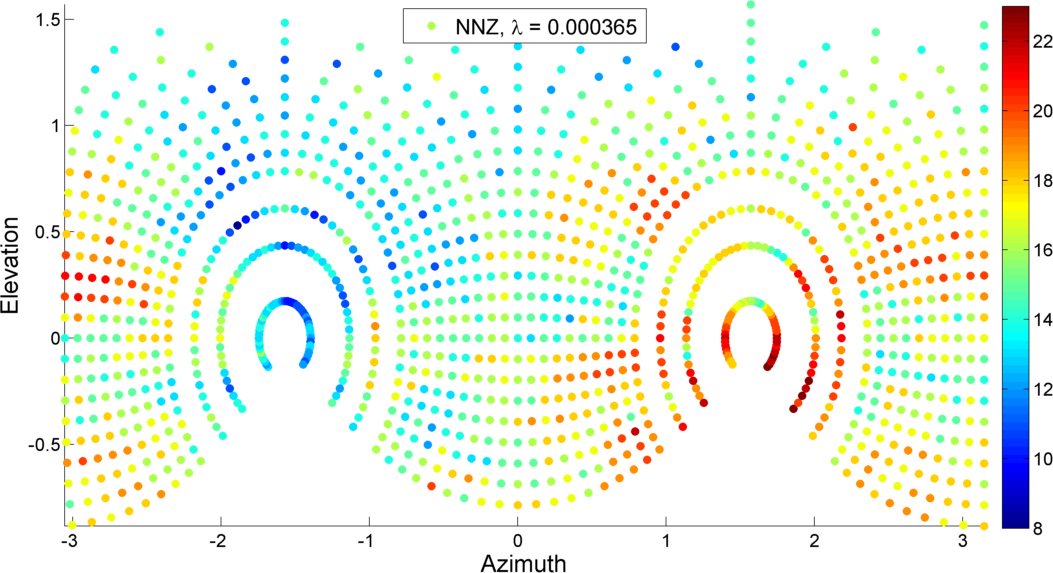}
\includegraphics[width=.49\textwidth]{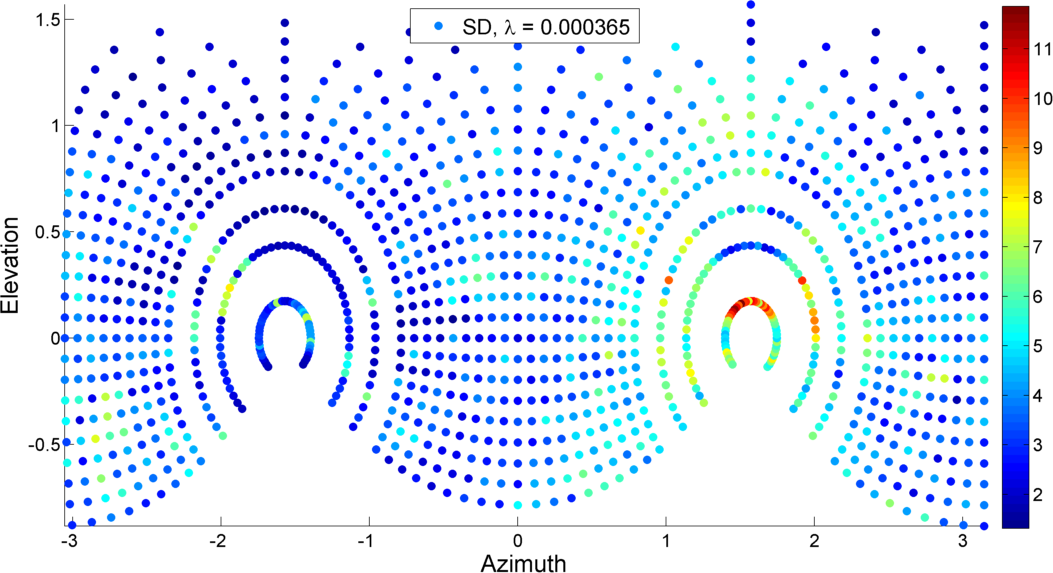}
\caption{A map of NNZE and SD error over the full spherical coordinate range for left-ear HRIR data. Note smaller NNZE / SD values on ipsilateral side.}
\label{FIG:NMF:SPARSE:SPHERE}
\end{figure}

Further analysis of the NNZE and of the SD error over the full set of HRIR measurement directions is shown in Fig. \ref{FIG:NMF:SPARSE:SPHERE}. Note that ipsilateral reflection filters have lower NNZE\footnote{The variability exhibited can not be due simply to total HRIR energy differences as they were all normalized during pre-processing.} and achieve lower SD error. This is understandable, as they do fit better into a ``resonance-plus-reflections'' model implied in this work. On the other hand, contralateral HRIR reconstruction requires larger NNZE and results in more distortion, presumably due to significant reflections occuring later than $K=25$ time samples; note that while some effects of head shadowing (attenuation / time delay) are removed in the preprocessing step, others may not be modeled accurately; on the other hand, accurate HRIR reproduction on contralateral side is not believed to be perceptually important \cite{LANGENDIJK}. Improvement in quality of contralateral HRIR reconstruction is a subject of future research. One approach is to learn separate HRIR decomposition, possibly with different length of $f$ / $g$ filters, for different sub-regions of space.

\begin{figure*}[ht]
  \centering
  \includegraphics[width=.49\textwidth]{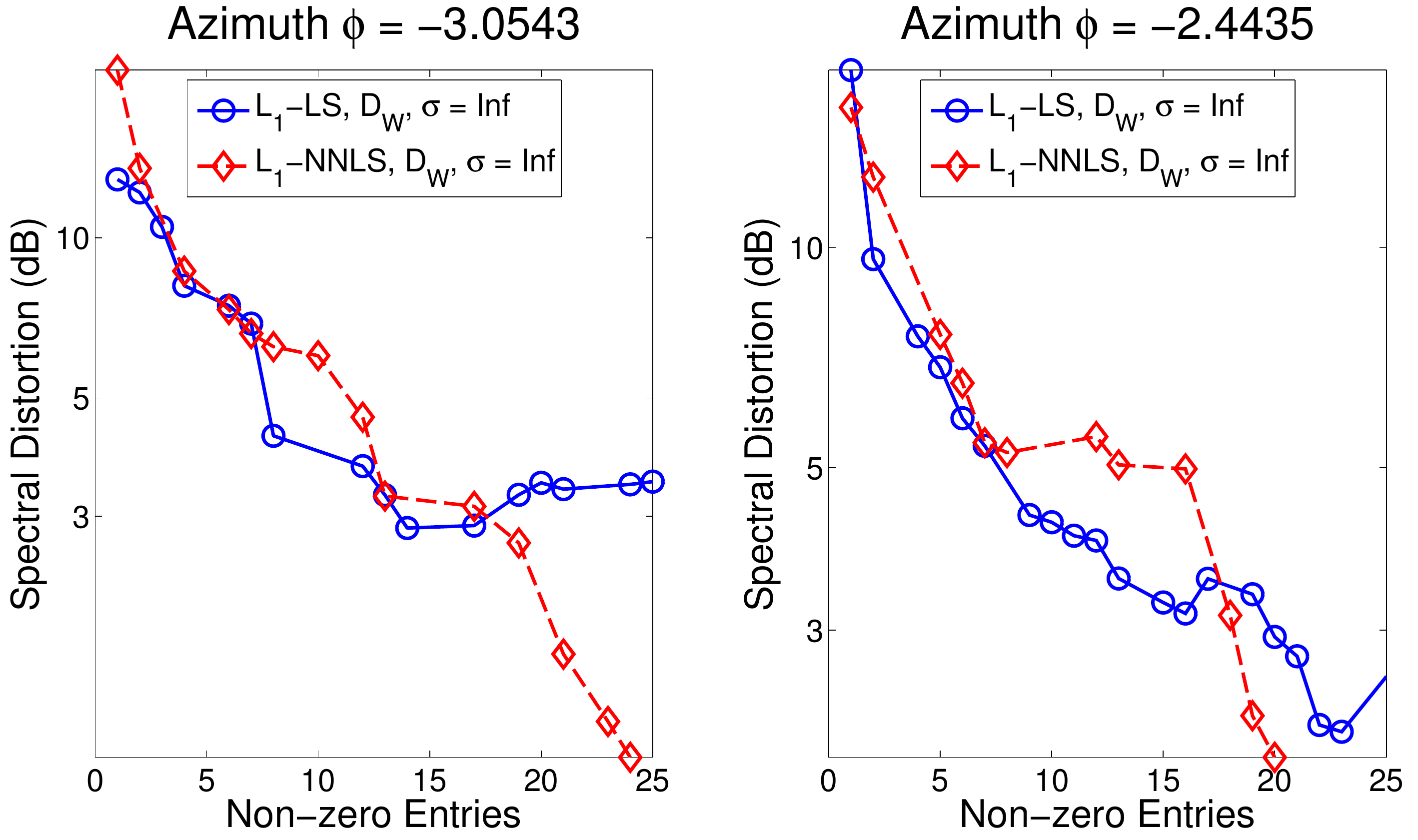}
    \includegraphics[width=.49\textwidth]{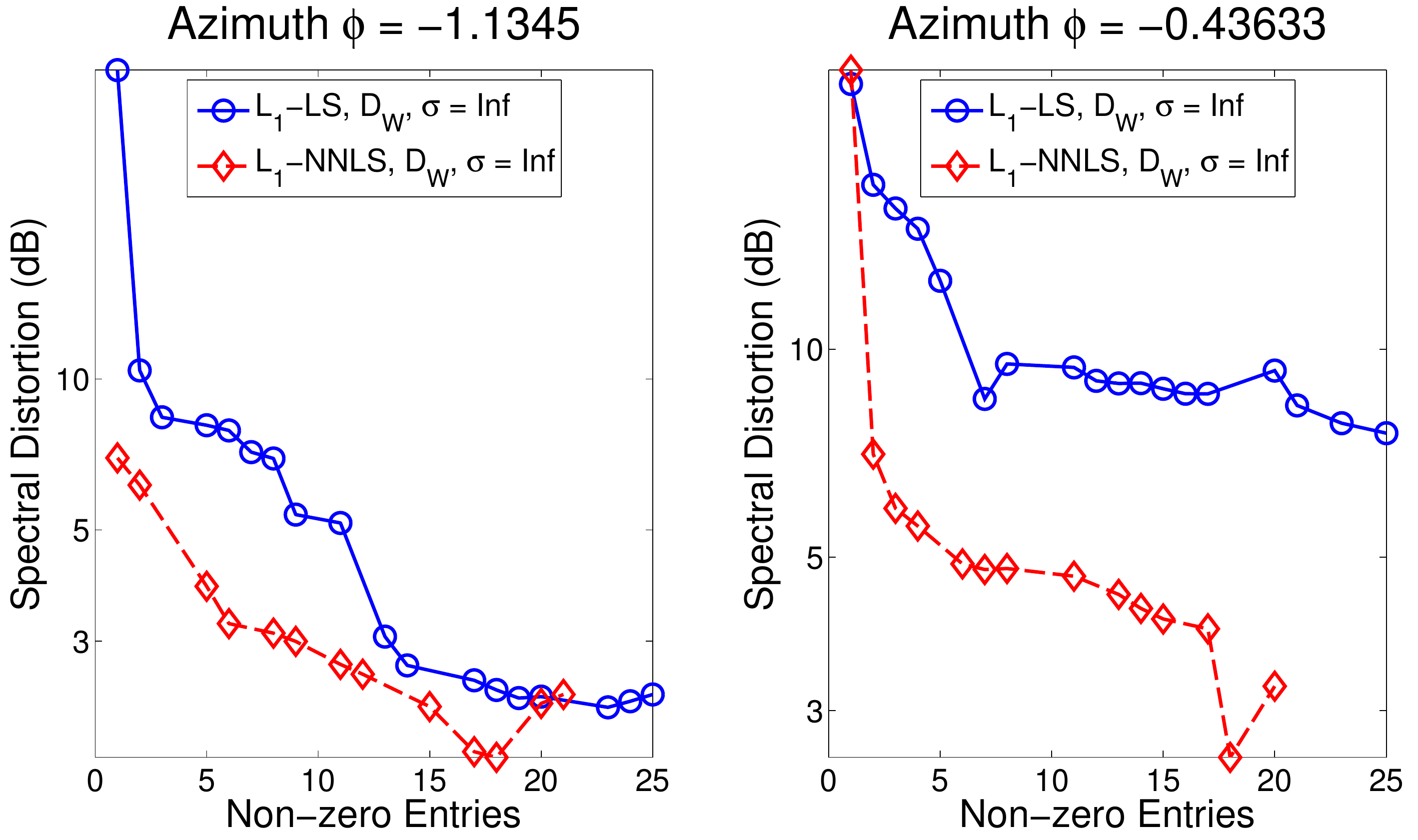}
  \includegraphics[width=.49\textwidth]{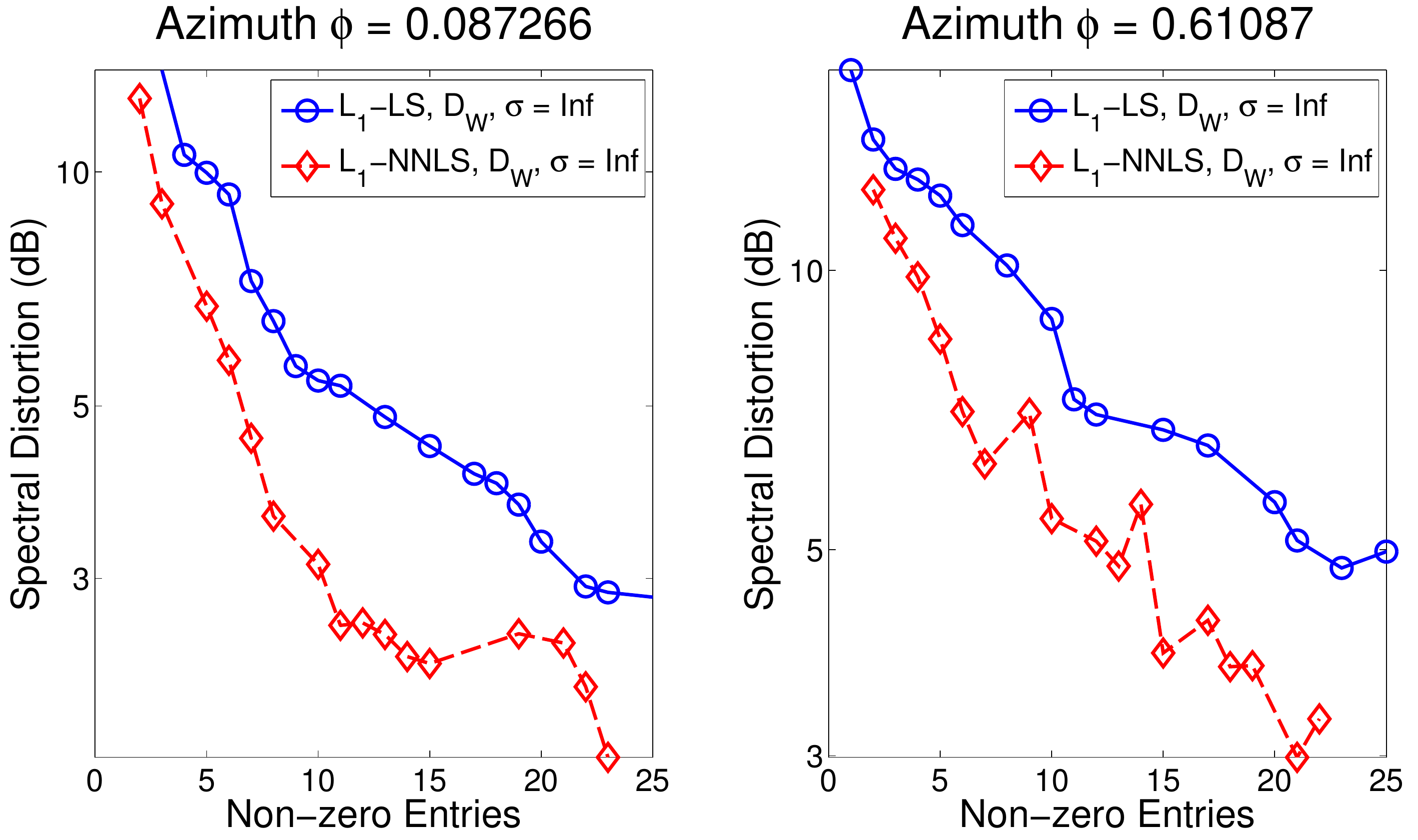}
    \includegraphics[width=.49\textwidth]{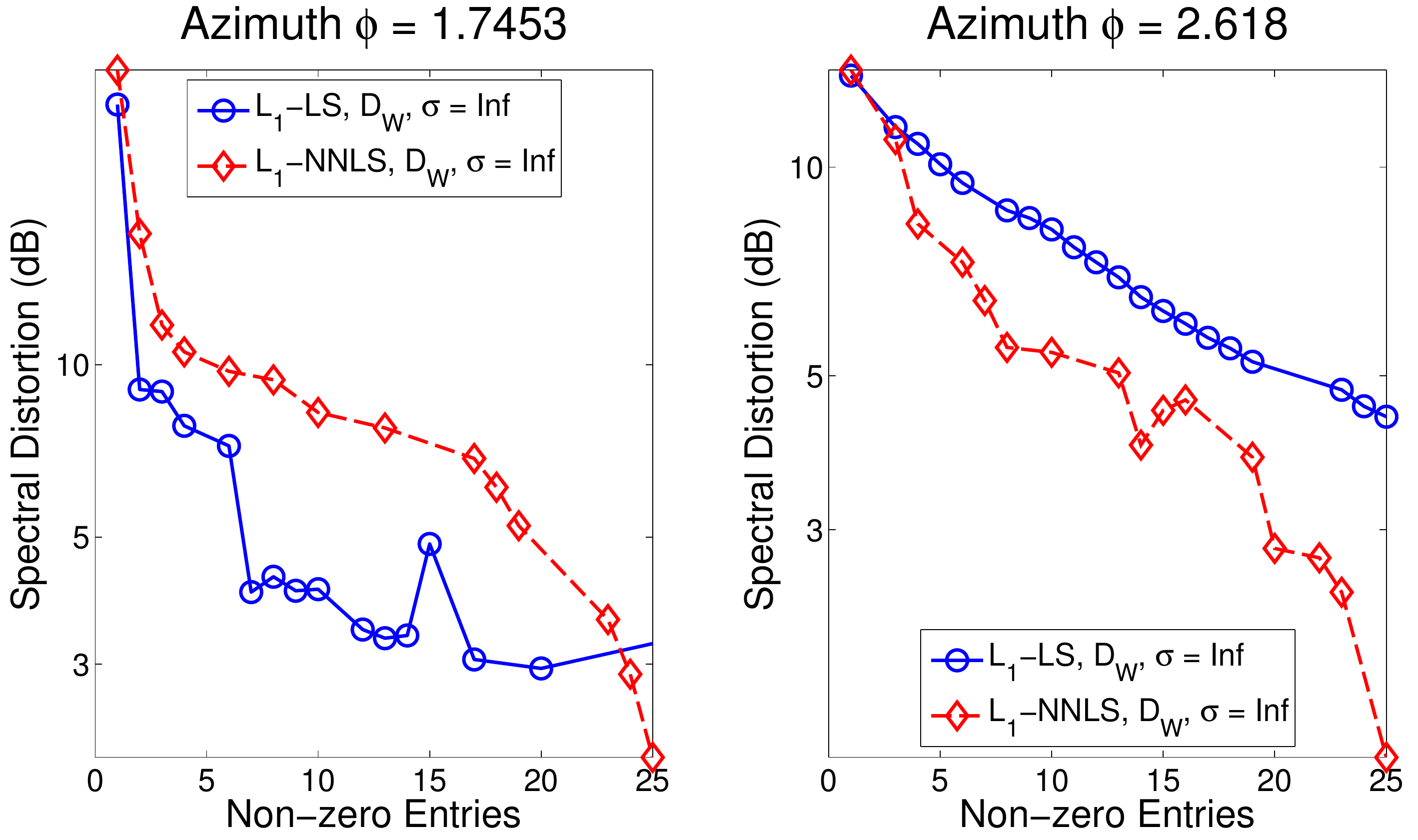} \\
      \includegraphics[width=.49\textwidth]{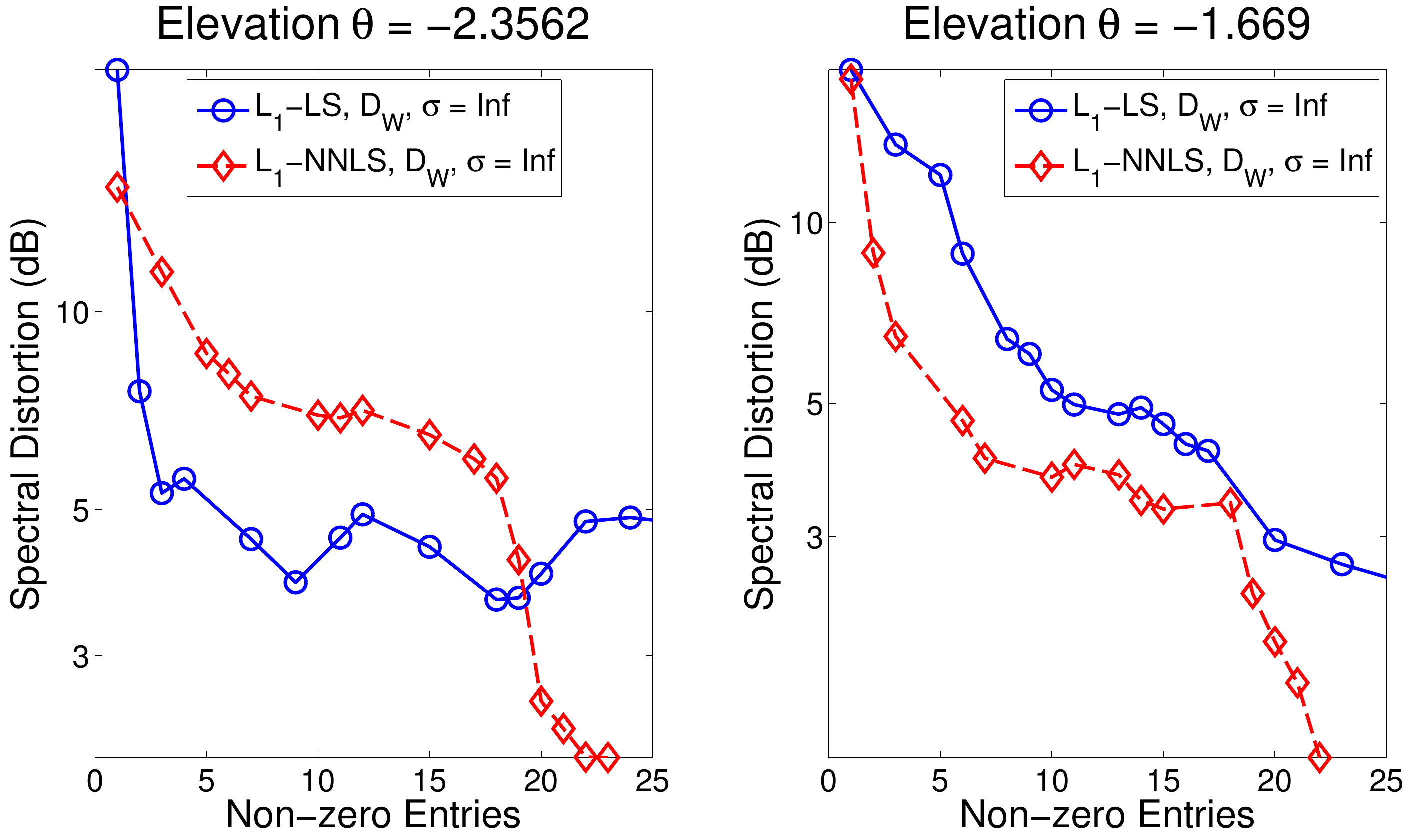}
    \includegraphics[width=.49\textwidth]{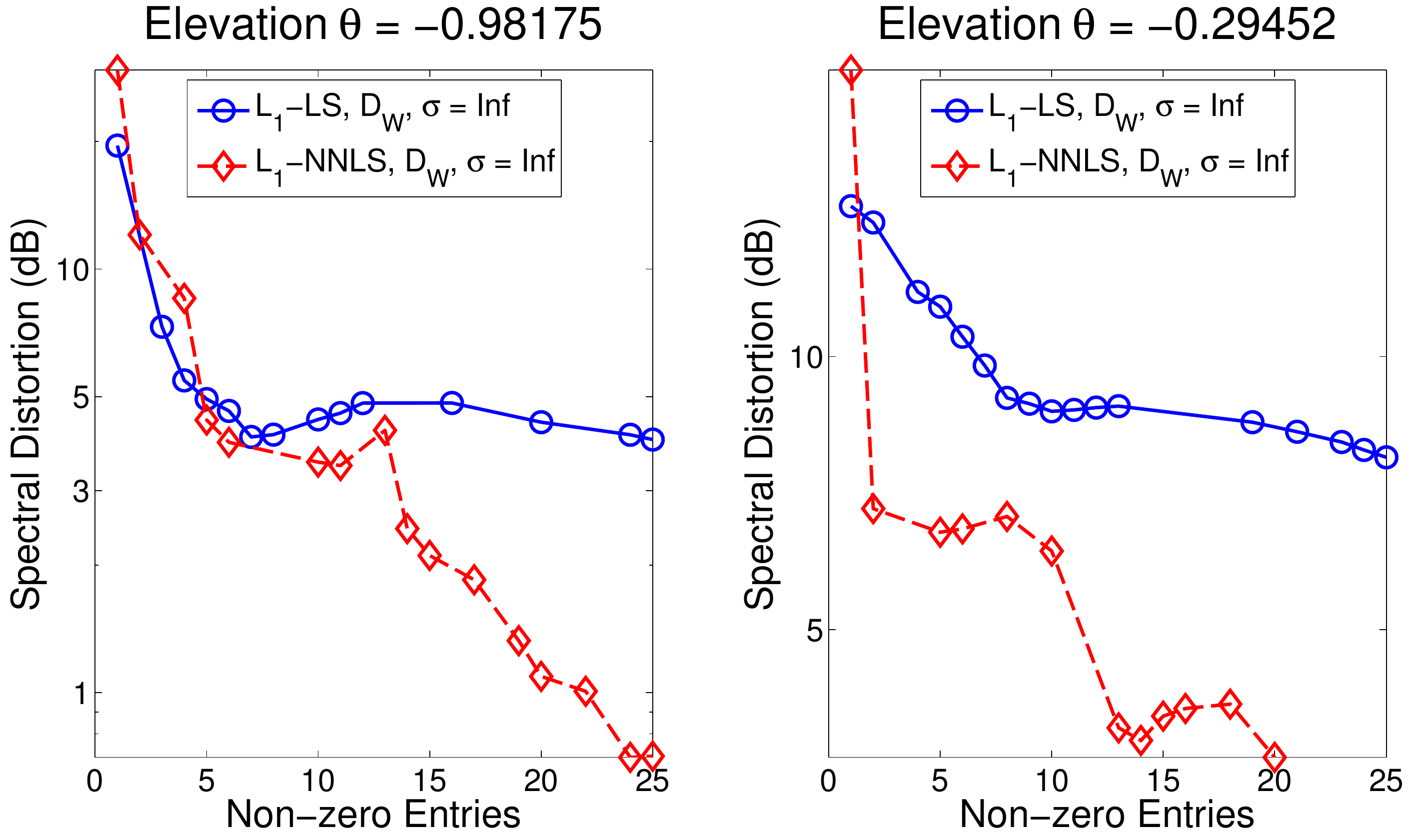}
  \includegraphics[width=.49\textwidth]{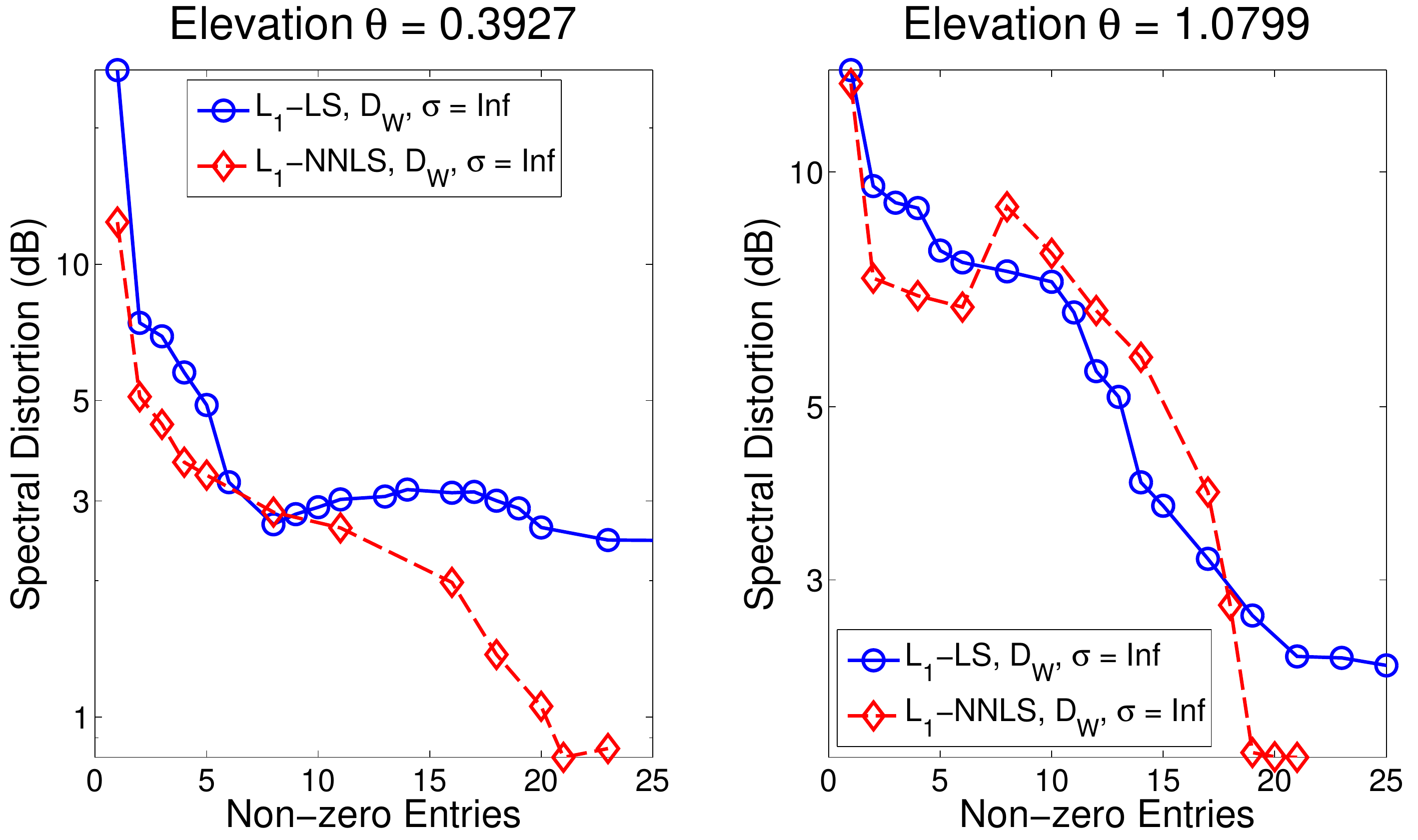}
    \includegraphics[width=.49\textwidth]{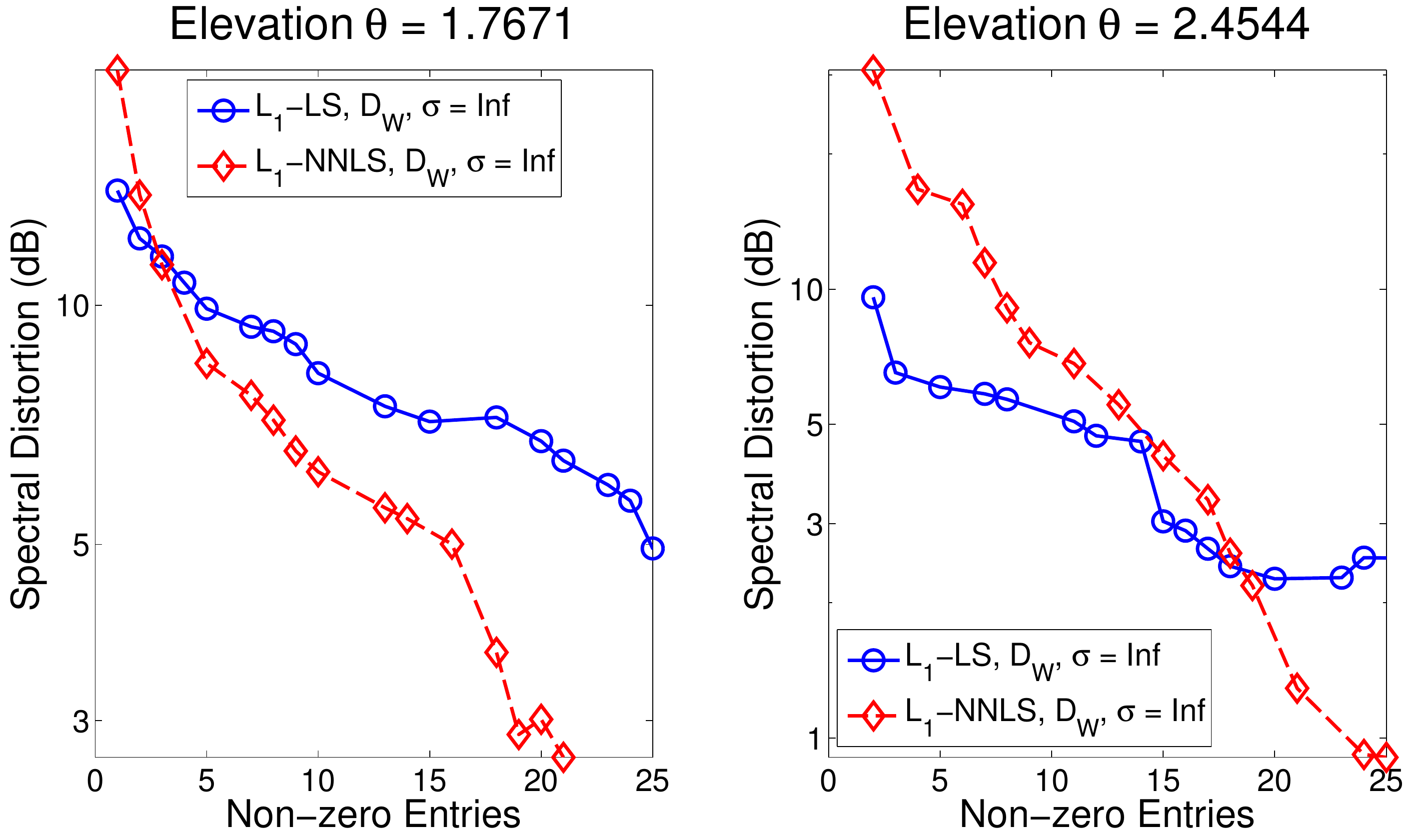} 
\caption{A comparison between varying-sparsity $L_1$-NNLS and $L_1$-LS solutions for selected directions on horizonal and median planes. Angles are listed in radians.}
\label{FIG:NMF:SPARSE:UNCON}
\end{figure*}

Finally, in Fig. \ref{FIG:NMF:SPARSE:UNCON} we compare the $L_1$-NNLS reconstruction against the naive $L_1$-LS reconstruction in terms of the convolution filter NNZE and SD error for varying $\lambda$ and a number of directions selected on horizontal and on medial planes. For all of these, the difference between solutions is less than $2.0$ dB SD; further, for 13 (out of 16) cases the $L_1$-NNLS solution has the same or better reconstruction error than naive $L_1$-LS solution in highly-sparse (NNZE $\leq K / 2$) case. This implies that our decomposition is able to find a resonance filter and a sparse set of early reflections that represent the HRTF better than the dominant magnitude components of the original HRIR \emph{per se}.

\subsection{Transformation Bandwidth Optimization}
\label{SEC:NMF:RESULTS:SIGMA}

Further reduction of the SD error is possible via use of transform functions defined in section \ref{SEC:NMF:SNTMF:SPARSE}. Application of these functions would result in different weights placed on different aspects of reconstructed HRIR. Hence, we investigate the selection of bandwidth term $\sigma$ in Eq. \ref{EQ:NMF:SNTMF:SPARSE:DW} with no $L_1$ penalty term ($\lambda = 0$) for the window transform\footnote{We omit the convolution transform $\mathcal{D}_C$ in experiments as applying a low-pass filter to the residuals entails a per-frequency error metric.}.

\begin{figure}[ht]
  \centering
\includegraphics[width=.49\textwidth]{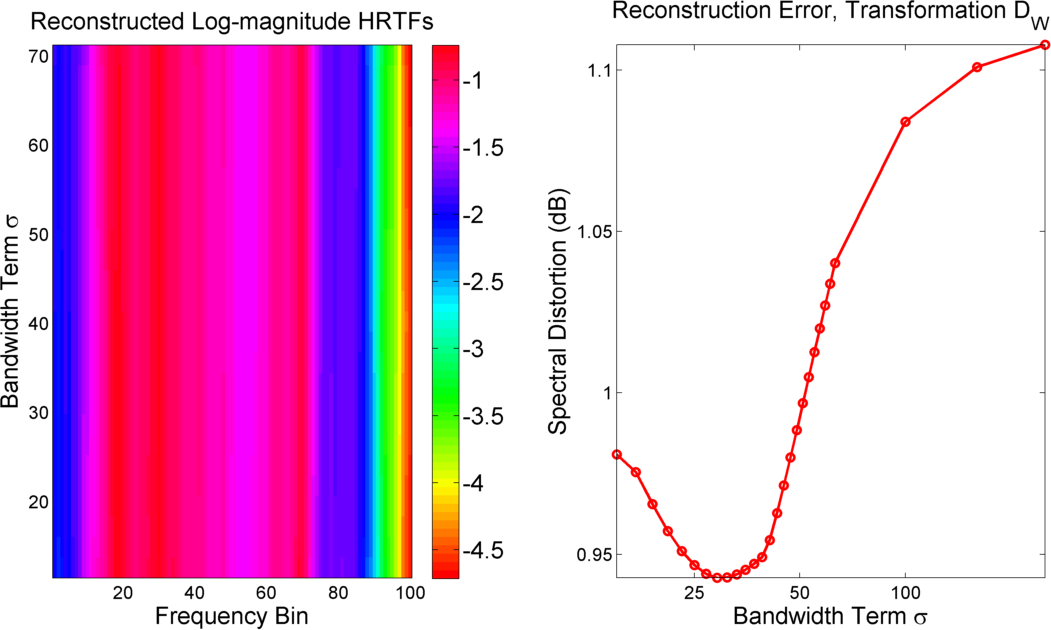} 
\caption{SD error dependence on bandwidth of window transform for a sample HRIR.}
\label{FIG:NMF:RESULTS:SIGMAW}
\end{figure}

As mentioned before, application of the window transform $\mathcal{D}_W$ causes smoothing in the frequency domain; the amount of smoothing depends on the bandwidth term $\sigma$. Fig. \ref{FIG:NMF:RESULTS:SIGMAW} shows the SD error dependence on $\sigma$ for one sample HRIR. Obviously as bandwidth $\sigma \rightarrow \infty$, the window transform becomes the identity transform; indeed, SD error stays constant for $\sigma > 70$. It can be seen though that the minimum SD error occurs at a finite $\sigma = 30$ (for this particular HRIR). The parameter $\sigma$ can be efficiently fine-tuned (via fast search methods) \emph{separately} for each HRIR in the subject's HRTF set. Table \ref{TAB:NMF:RESULTS:SIGMAW} compares the SD error obtained over the grid of $\sigma = [15 + ((0:24)*2), 100, 160, 250]$ using window transform to the SD error with identity transform (which is the same as window transform with $\sigma \rightarrow \infty$) across horizontal / median plane and over all HRTF set directions. It can be seen that on average, such tuning decreases the SD error by about $10\%$. 

\begin{table}
\caption{Mean spectral distortion for individually tuned $\mathcal{D}_{W,\sigma}$}
\centering
\begin{tabular}{|c|c|c|c|}
\hline
&\textbf{H-plane}&\textbf{M-plane}&\textbf{All directions}\\\hline
\textbf{$\sigma \rightarrow \infty$}& 2.72 & 1.73 &  2.49 \\\hline
\textbf{Tuned $\sigma$ }& 2.53 & 1.57 &  2.24 \\\hline
\end{tabular}
\label{TAB:NMF:RESULTS:SIGMAW}
\end{table}

\subsection{Computational Cost}

Consider the cost of computing the $i^{th}$ sample of $(x*y)_i$ where $*$ is the convolution operation. Direct time-domain convolution requires $\min \CBRAK{\ABS{x}, \ABS{y}}$ real floating-point operations, where $\ABS{x}, \ABS{y}$ is the NNZE in each filter. In practice, convolution is normally done in blocks of fixed size (so-called partitioned convolution). In case of time-domain processing, partitioned convolution incurs neither memory overhead nor latency.

At the same time, the state-of-the-art frequency-domain implementation \cite{JOHNSON} requires $\frac{68}{9}(\ABS{y} \log_2 \ABS{y} + \ABS{y}) / (\ABS{y} - \ABS{x} + 1)$ complex floating-point operations per output sample. For a long input signal (e.g. $|y| = 44100$ -- i.e. one second at CD audio quality), time-domain algorithm is faster than frequency-domain implementation for $|x| < 127$. Further, in real-time processing, latency becomes an issue, and one must use partitioned convolution (with reasonably small block size) and the \emph{overlap-and-save} algorithm \cite{OPPENHEIM}. In order to achieve e.g. $50$ ms latency, one must have $|y| = 2205$. For this segment length, direct time-domain convolution incurs less computational cost when $|x| < 90$. Thus, a time-domain convolution using sparse filter $x$ as derived in this paper is arguably quite beneficial to the computational load incurred by the VAD engine.

\section{Discussion}

While our study presents the theoretical derivation of our factorization algorithm, a number of practical concerns have been omitted for reasons of scope. We provide a number of remarks on these below.

First, an optimal NNZE is hardware dependent, as the crossover point between time-domain and frequency-domain convolution costs depends on the computational platform as well as on the specific implementations of both. For example, specialized digital signal processors can perform efficient real time-domain convolution via hardware delay lines whereas being less optimized for handling complex floating-point operations necessary for fast Fourier transform.

Second, the target reconstruction error can be adjusted to match a desired fidelity of spatialization. For instance, early reflections off nearby environmental features may have to be spatialized more distinctly than a number of low-magnitude later reflections that collectively form the reverberation tail. Further, the need to individually optimize the penalty term $\lambda$ for each direction depends also on desired sparsity (i.e. computational load) versus SD error trade-off. Such real-time load balancing is an open challenge that depends on available computational resources on specific hardware platform.

Certain obvious extensions of the work presented has also not been fully described for clarity. We note that using non-zero $\lambda$ term and varying the bandwidth $\sigma$ in $\mathcal{D}_W$, $\mathcal{D}_C$ transforms could lead to decrease in SD error at the same NNZE when tuned. A set of bandpass transformations that constitute the orthogonal basis for the discrete Fourier transform could also be used, as in this case the error could be weighted individually in each frequency band to match the listener's characteristics (e.g. by using the equal loudness contours in frequency).

Another consideration is the choice of the cost function in Eq. \ref{EQ:NMF:SNTMF:DEF}, which currently omits prior information on the HRIR measurement direction distribution. It may be undesirable to place equal weight on all directions if those are in fact spaced non-uniformly. Instead, the sample residual can be biased by introducing a kernel transformation $\mathcal{D} \in \field{R}^{N \times N}$ of the HRIR measurement directions ($\mathcal{D}_{ij}$ is a kernel function evaluation between directions $i^{th}$ and $j^{th}$) into the cost function $\trace{ (X-FG^T) \mathcal{D}^{-1} (X-FG^T)^T}$, which would decorrelate HRIR reconstruction error in densely-sampled area and thus avoid giving preferential treatment to these areas while optimizing.

\section{Conclusions}
\label{SEC:NMF:CONC}

We have presented a modified semi-NMF matrix factorization algorithm for Toeplitz constrained matrices. The factorization represent each HRIR in a collection as a convolution between a common ``resonance filter'' and specific ``reflection filter''. The resonance filter has mixed sign, is direction-independent, and is of length comparable to original HRIR length. The reflection filter is non-negative, direction-dependent, short, and sparse. The tradeoff between sparsity and approximation error can be tuned via the regularization parameter of $L_1$-NNLS solver, which also has the ability to place different weights on errors in different frequency bands (for HRTF) or at different time instants (for HRIR). Comparison between HRIR reconstructed using the proposed algorithm and $L_1$-LS reference solution shows that the former has much better sparsity-to-error tradeoff, thus allowing for high-fidelity latency-free spatial sound presentation at very low computational cost.


%
%


\ifCLASSOPTIONcaptionsoff
  \newpage
\fi



%

\bibliographystyle{IEEEtran}
\bibliography{masterbib}




\end{document}